\documentclass[10pt,twocolumn]{article}
\usepackage{cite}
\usepackage{amsmath,amssymb,amsfonts}
\usepackage{algpseudocode}
\usepackage{algorithm}
\usepackage{authblk}
\usepackage{mathtools}
\usepackage{hyperref}
\usepackage{xfrac}
\usepackage{diagbox}
\usepackage{graphicx}
\usepackage{textcomp}
\usepackage{xcolor}
\usepackage[cm]{fullpage}
\def\BibTeX{{\rm B\kern-.05em{\sc i\kern-.025em b}\kern-.08em
    T\kern-.1667em\lower.7ex\hbox{E}\kern-.125emX}}

\newcommand{\etal}{\textit{et al.}}

\hypersetup{
    colorlinks=true,
    linkcolor=blue,
    urlcolor=blue,
}

\begin{document}

\title{The Two-Pass Softmax Algorithm}
\date{}

\author[1,2]{Marat Dukhan\thanks{Corresponding Author: maratek@google.com}}
\author[1]{Artsiom Ablavatski}
\affil[1]{Google Research}
\affil[2]{Georgia Institute of Technology}
\setcounter{Maxaffil}{0}
\renewcommand\Affilfont{\itshape\small}

\setlength{\tabcolsep}{1.0em}

\maketitle

\begin{abstract}
The softmax (also called softargmax) function is widely used in machine learning models to normalize real-valued scores into a probability distribution. To avoid floating-point overflow, the softmax function is conventionally implemented in three passes: the first pass to compute the normalization constant, and two other passes to compute outputs from normalized inputs. We analyze two variants of the Three-Pass algorithm and demonstrate that in a well-optimized implementation on HPC-class processors performance of all three passes is limited by memory bandwidth.

We then present a novel algorithm for softmax computation in just two passes. The proposed Two-Pass algorithm avoids both numerical overflow and the extra normalization pass by employing an exotic representation for intermediate values, where each value is represented as a pair of floating-point numbers: one representing the ``mantissa'' and another representing the ``exponent''.

Performance evaluation demonstrates that on out-of-cache inputs on an Intel Skylake-X processor the new Two-Pass algorithm outperforms the traditional Three-Pass algorithm by up to 28\% in AVX512 implementation, and by up to 18\% in AVX2 implementation. The proposed Two-Pass algorithm also outperforms the traditional Three-Pass algorithm on Intel Broadwell and AMD Zen 2 processors.

To foster reproducibility, we released an open-source implementation of the new Two-Pass Softmax algorithm and other experiments in this paper as a part of XNNPACK library at \href{http://www.github.com/google/XNNPACK}{GitHub.com/google/XNNPACK}.

\end{abstract}

\section{Introduction}

The {\it softmax} (also called {\it softargmax}) function is a smooth approximation to the {\it argmax} function. The softmax function $\sigma(x)$ operates on a vector of real-valued scores $x_i$ and normalizes them into a probability distribution
$$p_i = \mathrm{\sigma}(x)_i = \frac{e^{x_i}}{\sum_k e^{x_k}}$$
where $p_i \ge 0$ and $\sum_i p_i = 1$.

The softmax function is commonly used in machine learning to give a probabilistic interpretation to outputs of classification models~\cite{bridle1990probabilistic}. Such machine learning models output a probability for every possible object class, and the number of classes in modern datasets can reach millions. For example, natural language processing models may predict probability distribution over each possible word in a vocabulary, and recommendation systems may model the probability distribution over users, products, web pages, or their interactions. Table~\ref{table:datasets} summarized number of classes on several public classification datasets.

\begin{table}[H]
\begin{center}
\begin{tabular}{|l|c|c|}
\hline
Dataset & Class description & Class Count \\
\hline\hline
ImageNet~\cite{deng2009imagenet} & Image category    & 21841 \\
One Billion Word~\cite{chelba2013one} & Unique Words & 793471 \\
Wikilinks~\cite{singh2012wikilinks} & Wikipedia pages & 2933659 \\
DepCC~\cite{depcc} & Web documents & 364.80 million \\
\hline
\end{tabular}
\end{center}
\caption{Characteristics of several public machine learning datasets.}
\label{table:datasets}
\end{table}

Hierarchical Softmax~\cite{goodman2001classes} (HSM) and its modifications~\cite{EfficientSoftmaxApproximationGPUs} are the common techniques to scale classification models to large number of classes. HSM models jointly consider the softmax function and the matrix-matrix multiplication that produced its input, and replace them by a low-rank approximation. Thus, HSM methods improve performance by reducing the matrix-matrix multiplication cost, and do so by approximating the original machine learning model.

Unlike previous research, we focus on the softmax function in the context of inference using a pre-trained model. This situation commonly arises in machine learning frameworks, such as TensorFlow~\cite{TensorFlow} or PyTorch~\cite{PyTorch}, when the training dataset or metaparameters needed to devise an accurate approximation to the model are not available. In this case, the model must be computed exactly according to the specification and unsafe approximations are not possible.

Our contributions in this paper are as follows:

\begin{itemize}
    \item We demonstrate that a well-optimized implementation of the softmax function can be memory-bound even in single-threaded execution. This result emphasizes the importance of eliminating memory operations for further improvements in the performance of the softmax function.
    \item We introduce a novel algorithm for computing the softmax function. The new algorithm employs an exotic representation for intermediate values, where each value is represented as a pair of floating-point numbers: one representing the ``mantissa'' and another representing the ``exponent''. Thanks to the special representation of intermediate results, the new algorithm needs only two passes over an input vector versus three passes for traditional algorithms.
    \item We present and evaluate high-performance implementations of the new Two-Pass softmax algorithms for the x86-64 processors with AVX2 and AVX512F SIMD extensions. The experimental study confirms the speedups of up to $28\%$ on an Intel Skylake-X system.
\end{itemize}

The proposed improvements to the softmax implementation are orthogonal to matrix-matrix multiplication optimizations, and can be combined with sparsification~\cite{SCNN, HolisticSparseCNN}, low-rank decomposition~\cite{Denton2014}, low-precision arithmetic~\cite{GEMMLOWP, FBGEMM, Tulloch2017}, or hardware acceleration~\cite{EIE, TPU} for the matrix-matrix multiplication that produce softmax input.

\section{Related work}

The softmax function is widely used in the modern machine learning models and algorithms. In particular, the softmax function gained wide popularity in Natural Language Processing as a building block for language models, which predict a word or n-gram out of a large vocabulary~\cite{LMLimits}. As the softmax function plays an important role in the machine learning models the several approximations have been proposed~\cite{HardwareSoftmax,HighSpeedLowComplexitySoftmax,SoftmaxVLSI,SVDSoftmax,SigSoftmax,EfficientHardwareSoftmax,EfficientSoftmaxApproximationGPUs} in order to make the calculations efficient and fast. We refer the reader to Jozefowicz~\etal~\cite{EfficientSoftmaxApproximationGPUs} for a good overview of the recent approaches to speed up the softmax calculations.

Although a lot work has been done to address the computation complexity of the softmax function on a large inputs, the proposed solutions either 1) require special hardware (e.g FPGA) or 2) require to use an approximation to the softmax function during model training. In the latter case the softmax approximation can not be directly applied to a pre-trained machine learning model, which dramatically limits its use in existing machine learning frameworks e.g. TensorFlow~\cite{TensorFlow} and PyTorch\cite{PyTorch}. Therefore, we developed a novel algorithm that improves performance of the original softmax function on widely available hardware, can be used with pre-trained machine learning models, and can be implemented in any machine learning framework.

\section{The Three-Pass Algorithm}
\label{sec:three_pass}

Direct calculation of the softmax function according to the formula $\mathrm{\sigma}(x)_i = \frac{e^{x_i}}{\sum_k e^{x_k}}$ is conjugate with numerical issues. Single-precision $e^{x}$ function overflows\footnote{Produce floating-point infinity because result is too large to be represented as a finite single-precision floating-point number} for $x > 89$ and underflows\footnote{Produce 0 because result is too small to be distinguishable from zero in the single-precision floating-point format} for $x < -104$, and, in turn, cause NaN\footnote{Not a Number: a special floating-point value defined by IEEE 754 standard indicating invalid result} outputs in the na\"ive implementations of the softmax function. In practice, the parts of the machine learning models that produce input to the softmax function are rarely bounded, and thus implementation can't assume that the input would fall into such narrow range.

In order to overcome numerical instability issues machine learning frameworks adapt a workaround by utilizing the equivalence~\cite{goodfellow2016deep}:

$$\mathrm{\sigma}(x)_i = \frac{e^{x_i}}{\sum_k e^{x_k}} = \frac{e^{\left (x_i - c\right )}}{\sum_k e^{\left (x_k - c\right )}}$$
which holds for any $c$ value. In particular, if $c = \underset{i}{\max {x_i}}$, then:
\begin{itemize}
    \item No inputs to $e^x$ function exceed zero
    \item There is at least one zero input to the $e^x$ function, and thus the denominator of the fraction is non-zero.
\end{itemize}
Together, these properties result in good numerical behavior of the computation and lead to Algorithm~\ref{algo:three_pass}.

\begin{algorithm}[H]
    \begin{algorithmic}
        \Function {SoftmaxThreePassRecompute}{$X$, $Y$}
            \State $N \gets \Call{Length}{X}$
            \State $\mu \gets \underset{1 \leq i \leq N}{\max} {X_i}$ \Comment{Pass 1: read X}
            \State $\sigma \gets \underset{1 \leq i \leq N}{\sum}\Call{Exp}{X_i - \mu}$
            \Comment{Pass 2: read X}
            \State $\lambda \gets \sfrac{1}{\sigma}$
            \ForAll{$1 \leq i \leq N$}
                \State $Y_i \gets \lambda \cdot \Call{Exp}{X_i - \mu}$
                \Comment{Pass 3: read X, write Y}
            \EndFor
        \EndFunction
    \end{algorithmic}
    \caption{The Three-Pass algorithm with re-computation of exponential function}
    \label{algo:three_pass}
\end{algorithm}

Both of the Pass 2 and the Pass 3 in Algorithm~\ref{algo:three_pass} compute $e^{\left ( x_i - \mu \right)}$ with the same $x_i$ and $\mu$ values. This observation hints a potential optimization: if computing $e^x$ function is expensive, we could save the computed $e^{\left ( x_i - \mu \right)}$ values to avoid recomputing them in the Pass 3. Such modification is presented in Algorithm~\ref{algo:three_pass_with_reloading}.

\begin{algorithm}[H]
    \begin{algorithmic}
        \Function {SoftmaxThreePassReload}{$X$, $Y$}
            \State $N \gets \Call{Length}{X}$
            \State $\mu \gets \underset{1 \leq i \leq N}{\max} {X_i}$ \Comment{Pass 1: read X}
            \State $\sigma \gets 0$
            \ForAll{$1 \leq i \leq N$}
                \State $Y_i \gets \Call{Exp}{X_i - \mu}$
                \Comment{Pass 2: read X, write Y}
                \State $\sigma \gets \sigma + Y_i$
            \EndFor
            \State $\lambda \gets \sfrac{1}{\sigma}$
            \ForAll{$1 \leq i \leq N$}
                \State $Y_i \gets \lambda \cdot Y_i$
                \Comment{Pass 3: read Y, write Y}
            \EndFor
        \EndFunction
    \end{algorithmic}
    \caption{The Three-Pass algorithm with re-loading of exponential computations}
    \label{algo:three_pass_with_reloading}
\end{algorithm}

Algorithm~\ref{algo:three_pass_with_reloading} computes $e^{\left ( x_i - \mu \right)}$ values only once, but this reduction in the number of computations comes at a cost: the second pass of Algorithm~\ref{algo:three_pass_with_reloading} does both a read and a write for each element, unlike Algorithm~\ref{algo:three_pass} where the second pass does only reads.

\section{The Two-Pass Algorithm}
\label{sec:two_pass}

The Three-Pass Algorithms~\ref{algo:three_pass} and~\ref{algo:three_pass_with_reloading} avoid numerical issues by normalizing inputs relative to their maximum value, but then require an additional memory pass to find the maximum value. In this section we suggest that it is possible to get the numerical stability without the extra memory pass, and present a practical Two-Pass algorithm for softmax computation.

The immediate reason for the numerical instability of a na\"ive softmax implementation is the saturation of $e^x$ function for inputs outside of the narrow range of $[-104, 89]$. Therefore, we have to look inside the $e^x$ function for a solution.

The $e^x$ function can be implemented in infinite number of ways, but practical implementations~\cite{AccurateTables,XScaleFP,IA64ElementaryFunctions,dukhan2013methods} in IEEE floating-point arithmetic follow the traditional structure of elementary function implementation~\cite{muller2010handbook}, and include three steps:

\begin{enumerate}
    \item \textbf{Range reduction}, where the problem of approximating $e^x$ on the infinite input domain $x \in (-\infty, +\infty)$ is reduced to a problem of approximating $e^x$ on a small finite range. For $e^x$, a natural range reduction derives from the equivalence
    $$e^x = e^{\overbrace{x - \log{2} \cdot \lfloor x \cdot \log_2{e}\rceil}^{t \in \left [ -\frac{\log 2}{2}, \frac{\log 2}{2} \right ]}} \cdot 2^{\overbrace{\lfloor x \cdot \log_2{e}\rceil}^{n \in \mathbb{Z}}}$$
    which decompose approximating $e^x$ on $x \in (-\infty, +\infty)$ into approximating $e^x$ on $t \in \left [ -\frac{\log 2}{2}, \frac{\log 2}{2} \right ]$ and multiplying the result by $2^n$ where $n$ is an integer.
    \item \textbf{Approximation} of the function on the reduced range, i.e. on $\left [ -\frac{\log 2}{2}, \frac{\log 2}{2} \right ]$ for $e^x$. This step is achieved through a polynomial or rational approximation, often in combination with table look-ups.
    \item \textbf{Reconstruction} of the final $e^x$ value from the approximation on the reduced range. For $e^x$, the reconstruction step consists of multiplication of $e^t$ by $2^n$, and can be achieved at low cost by manipulating the exponent field of a binary floating-point number $m \coloneqq e^t$. \textbf{It is this step where the underflow and overflow situations arise}: $e^t \in \left [ \frac{\sqrt 2}{2}, \sqrt 2 \right ]$ and thus always fits into a single-precision floating-point number, but $n = \lfloor x \cdot \log_2{e}\rceil$ can exceed the range of the exponent field, causing underflow or overflow.
\end{enumerate}

The key idea that enables the Two-Pass Softmax algorithm is to remove the reconstruction step, and instead keep the result of $e^{x}$ as a pair of floating-point values $(m, n)$, where $m = e^t$. Mathematically, $e^x = m \cdot 2^n$, but in general this expression can not be evaluated in floating-point arithmetic without overflowing or underflowing the result. The representation of $n$ as a \textit{floating-point} number is important: although $n$ is by design always an integer, it can have a very large magnitude, and fall outside of the range of standard integer formats. Therefore, the result $e^x$ must be represented as two, rather than one, floating-point numbers. Using multiple floating-point numbers to represent the real-valued result has a long history in double-double, triple-double, quad-double~\cite{hida2001algorithms} representations and Error-Free Transformations~\cite{graillat2007error}. However, these representations use multiple floating-point numbers to improve precision of floating-point arithmetic, whereas we suggest to use two floating-point numbers to extend its dynamic range.

\begin{algorithm}[H]
    \begin{algorithmic}
        \Function {SoftmaxTwoPass}{$X$, $Y$}
            \State $N \gets \Call{Length}{X}$
            \State $m_{sum} \gets 0$
            \State $n_{sum} \gets -\infty$
            \ForAll{$1 \leq i \leq N$}
                \State $m_i, n_i \gets \Call{ExtExp}{X_i}$
                \Comment{Pass 1: read X}
                \State $n_{max} \gets \Call{Max}{n_i, n_{sum}}$
                \State $m_{sum} \gets m_i \cdot 2^{n_i - n_{max}} + m_{sum} \cdot 2^{n_{sum} - n_{max}}$
                \State $n_{sum} \gets n_{max}$
            \EndFor
            \State $\lambda_{sum} \gets \sfrac{1}{m_{sum}}$
            \ForAll{$1 \leq i \leq N$}
                \State $m_i, n_i \gets \Call{ExtExp}{X_i}$
                \Comment{Pass 2: read X, write Y}
                \State $Y_i \gets m_i \cdot \lambda_{sum} \cdot 2^{n_i - n_{sum}}$
            \EndFor
        \EndFunction
    \end{algorithmic}
    \caption{The Two-Pass softmax algorithm. $\textrm{ExtExp}$ denotes an exponential function that produce a pair $(m, n)$ of floating-point values.}
    \label{algo:two-pass}
\end{algorithm}

Algorithm~\ref{algo:two-pass} presents the softmax computation in just two passes by implementing addition for $(m, n)$ representation. The reduction pass keeps track of the running maximum $n$ value among all elements, and accumulates the scaled $m$ values to the running sum. It avoids the floating-point overflow by scaling $m$ values by the difference between the corresponding $n$ values and maximum of $n$ values. As this difference is never positive, $m$ values are never scaled up, which ensures the absence of the floating-point overflow.

\section{Theoretical Analysis}

\begin{table*}
    \begin{center}
        \begin{tabular}{|l|c|c|c|c|}
            \hline
            Algorithm & Memory reads & Memory writes & Bandwidth cost \\
            \hline\hline
            Three-Pass (Recompute)~\ref{algo:three_pass} & 3N & \textbf{1N} & 4N \\
            Three-Pass (Reload)~\ref{algo:three_pass_with_reloading} & 3N & 2N & 5N \\
            Two-Pass~\ref{algo:two-pass} & \textbf{2N} & \textbf{1N} & \textbf{3N} \\
            \hline
        \end{tabular}
    \end{center}
    \caption{Theoretical analysis of memory complexity and bandwidth costs of the three softmax algorithms.}
    \label{table:analysis}
\end{table*}

While the number of memory passes in the presented softmax algorithms is evident from the names, the number of actual memory operations is more nuanced. Every pass of the Three-Pass algorithm with Recomputing reads the input array, while the last pass also writes the output array. The Three-Pass algorithm with Reloading just reads the input array in the first pass, reads the input array and writes the output array in the second pass, and reads-modifies-writes the output array in the last pass. The Two-Pass algorithm reads the input array in both passes, and also writes the output array in the second pass. Thus, the memory bandwidth requirements of the Two-Pass algorithm are similar to just the last two passes of the Three-Pass algorithm with Recomputing.

Table~\ref{table:analysis} summarize the number of memory reads, memory writes, and the memory bandwidth cost for the three algorithms on arrays of $N$ elements. Per Table~\ref{table:analysis}, the Two-Pass algorithm has a memory bandwidth advantage of $33\%$ over the Three-Pass algorithm with Recomputing and $67\%$ over the Three-Pass algorithm with Reloading. In practice, we should treat these numbers as upper bounds, because higher computational complexity of the Two-Pass algorithm cuts into gains from bandwidth reduction.

\section{Experimental Evaluation}

\subsection{Platform} We evaluate the performance of the three softmax algorithms on the Intel Xeon W-2135 processor based on Skylake-X microarchitecture and with the characteristics listed in Table~\ref{table:processor}. To improve performance stability we disabled dynamic frequency scaling in the processor for the duration of our experiments.

\begin{table}[H]
    \begin{center}
        \begin{tabular}{|l|c|}
            \hline
            Characteristic & Value \\
            \hline\hline
            Microarchitecture & Skylake-X \\
            Number of cores & 6 \\
            Number of hyperthreads & 12 \\
            Base frequency & 3.7 GHz \\
            L1 cache size (per core) & 32 KB \\
            L2 cache size (per core) & 1 MB \\
            L3 cache size (shared by all cores) & 8.25 MB \\
            AVX2 / AVX512 FMA throughput & 2 / cycle \\
            AVX2 / AVX512 FMA latency & 4 cycles \\
            \hline
        \end{tabular}
    \end{center}
    \caption{Characteristics of the Intel Xeon W-2135 processor used for experimental evaluation of the softmax algorithms.}
    \label{table:processor}
\end{table}

Additionally, in Sec.~\ref{sec:alternative_processors} we replicate a subset of experiments on a Broadwell-based Intel Xeon E5-2696 v4 processor and on an AMD Ryzen 9 3900X processor with Zen 2 microarchitecture.

\subsection{Protocol} We use Google Benchmark framework to estimate sustained performance of the softmax implementations. We set the minimum run-time for each measurement to 5 seconds, repeat each measurement 25 times and record the median of the 25 runs. In each benchmark run we simulate the cache state during neural network inference: output vector is evicted from the cache before each iteration, but input tensor stays in cache as long as it fits.

\subsection{Implementation} We developed highly optimized implementations of the Three-Pass Algorithms~\ref{algo:three_pass} and~\ref{algo:three_pass_with_reloading}, and the Two-Pass Algorithm~\ref{algo:two-pass} in C. For all three algorithms, we did two templated implementations targeting AVX2 and AVX512 instruction sets. We expressed high-level optimization parameters, such as unroll factor for the loops and the number of accumulator variables in reduction functions, as meta-parameters of the templated implementations, and employed auto-tuning to discover their optimal values.

An efficient implementation of vectorized $e^x$ function is a key component of all softmax variants. For our implementation, we adapted the throughput-optimized methods of Dukhan and Vuduc~\cite{dukhan2013methods} to single-precision floating-point evaluation. Algorithm~\ref{algo:exp} presents the resulting table-free, branch-free, and division-free algorithm.

\begin{algorithm}[H]
    \begin{algorithmic}
        \Function {Exp}{$x$}
            \State $n \gets \lfloor x \cdot \log_{2}{e} \rceil$
            \State $t \gets x - n \cdot \log 2$
            \Comment{Cody-Waite range reduction}
            \State $p \gets 1 + t(c_1 + t (c_2 + t (c_3 + t (c_4 + t \cdot c_5)))))$
            \Comment{Polynomial approximation}
            \State $y \gets p \cdot 2^n$
            \Comment{Reconstruction}
            \State \textbf{return} $y$
        \EndFunction
    \end{algorithmic}
    \caption{Calculation of \textbf{$e^x$} in the Three-pass softmax algorithms}
    \label{algo:exp}
\end{algorithm}

Algorithm~\ref{algo:exp} follows the traditional structure of elementary function implementation~\cite{muller2010handbook}, described in Sec.~\ref{sec:two_pass}. It starts with a range reduction to reduce approximation on the infinite input domain to approximation on a small and finite range. The calculation of the reduced argument $t$ in Algorithm~\ref{algo:exp} uses Cody-Waite range reduction~\cite{cody1980software}: $\log{2}$ is represented as a sum of two single-precision constants, $log_{hi}{2}$ and $log_{lo}{2}$, to improve the accuracy of this step. Range reduction results in a reduced argument $t$ in the $[-\frac{\log 2}{2}, \frac{\log 2}{2}]$ range and a reduced integer argument $n$. Next, $e^{t}$ is approximated on $[-\frac{\log 2}{2}, \frac{\log 2}{2}]$ with a degree-5 polynomial. The polynomial coefficients are produced by the algorithm of Brisebarre and Chevillard~\cite{brisebarre2007efficient} as implemented in Sollya software~\cite{chevillard2010sollya}. Following~\cite{dukhan2013methods}, we evaluate the approximation polynomial with Horner scheme using Fused Multiply-Add instructions to minimize the number of floating-point instructions and maximize the throughput. In the last stage the Algorithm~\ref{algo:exp} reconstructs the final output value of the function by multiplying polynomial approximation result $p$ by $2^n$. In AVX2 implementation, we do this multiplication by directly manipulating floating-point exponent to construct a scale number $s$:
\[ s \coloneqq
   \begin{cases} 
      2^n & n >= -126 0 \\
      0 & n < 126\leq x 
   \end{cases}
\]
and compute the final step as $y \gets p \cdot s$. This reconstruction trick has two buit-in assumptions: the argument $x$ to the $e^x$ is negative\footnote{Always the case for the $e^x$ evaluation in the Three-Pass softmax algorithms.}, and subnormal floating-point numbers can be flushed to zero without significant accuracy impact.
The reconstruction step in the AVX512 implementation leverage the new \textrm{VSCALEFPS} instruction~\cite{cornea2015intel} which computes $p \cdot 2^n$ as a single hardware operation.

The resulting $e^x$ implementation has maximum error under 2 ULP, validated through exhaustive evaluation on all valid inputs. This accuracy is comparable to other vectorized elementary function implementations, e.g. SIMD functions in GNU LibM library guarantee maximum error under 4 ULP.

Implementation of the $\textrm{ExtExp}$ in the Two-Pass softmax algorithm is similar to Algorithm~\ref{algo:exp} with the reconstruction step removed. Thus, implementations of both the Three-Pass and the Two-Pass algorithms use exactly the same range reduction and approximating polynomials to compute the exponential function.

\subsection{The Three-Pass Algorithms and Bandwidth Saturation}

\label{sec:three_pass_evaluation}

\begin{figure}[H]
    \includegraphics[width=\linewidth]{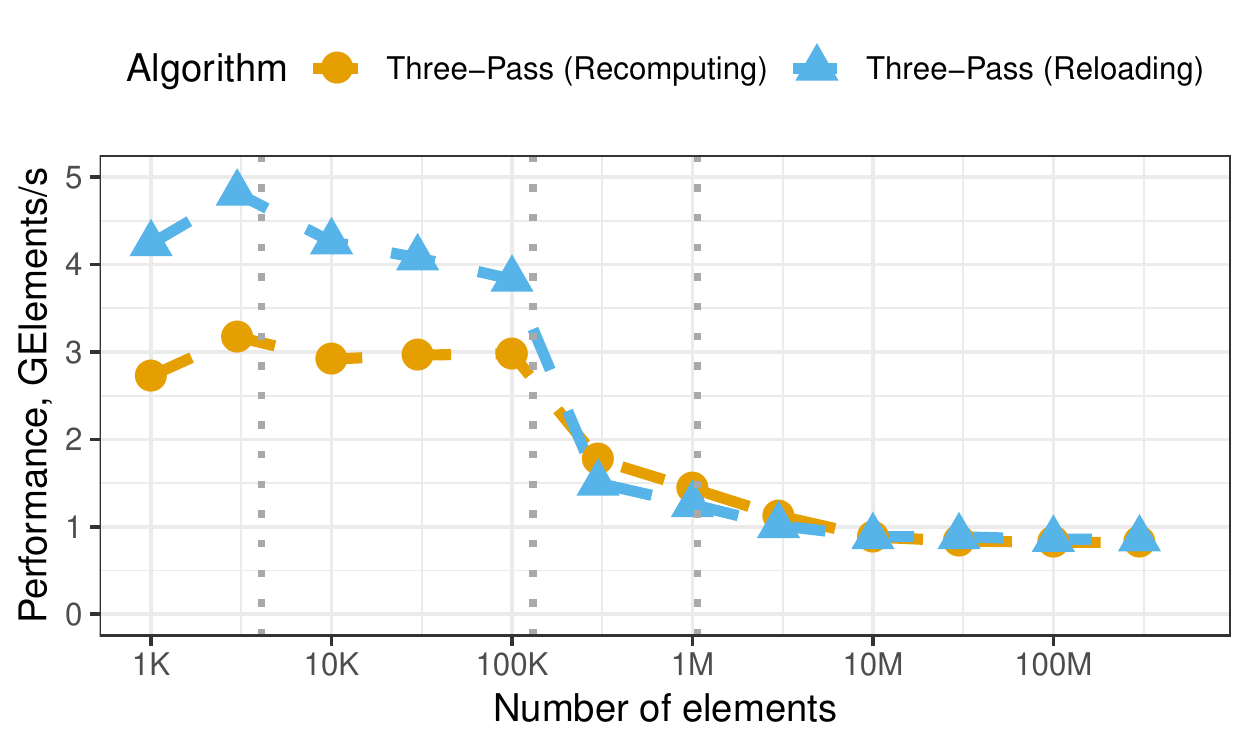}
    \caption{Performance comparison of the Softmax algorithms~\ref{algo:three_pass} and~\ref{algo:three_pass_with_reloading} in the AVX512 implementations on the Skylake-X system. Gray dotted lines denote boundaries of level-1, level-2, and level-3 caches.}
    \label{fig:avx512-threepass}
\end{figure}

Fig.~\ref{fig:avx512-threepass} presents the performance of the Three-Pass softmax Algorithm~\ref{algo:three_pass} with recomputing of exponentials and the Three-Pass softmax Algorithm~\ref{algo:three_pass_with_reloading} with reloading of computed exponentials in the AVX512 implementations. Reloading of exponential computations delivers $30-55\%$ speedup when the data is small enough to fit into private L1 and L2 caches, but turns into $15\%$ slowdown when operating on L3 cache, and eventually levels off at $4-6\%$ advantage when working set exceeds last-level cache.

\begin{figure}[H]
    \includegraphics[width=\linewidth]{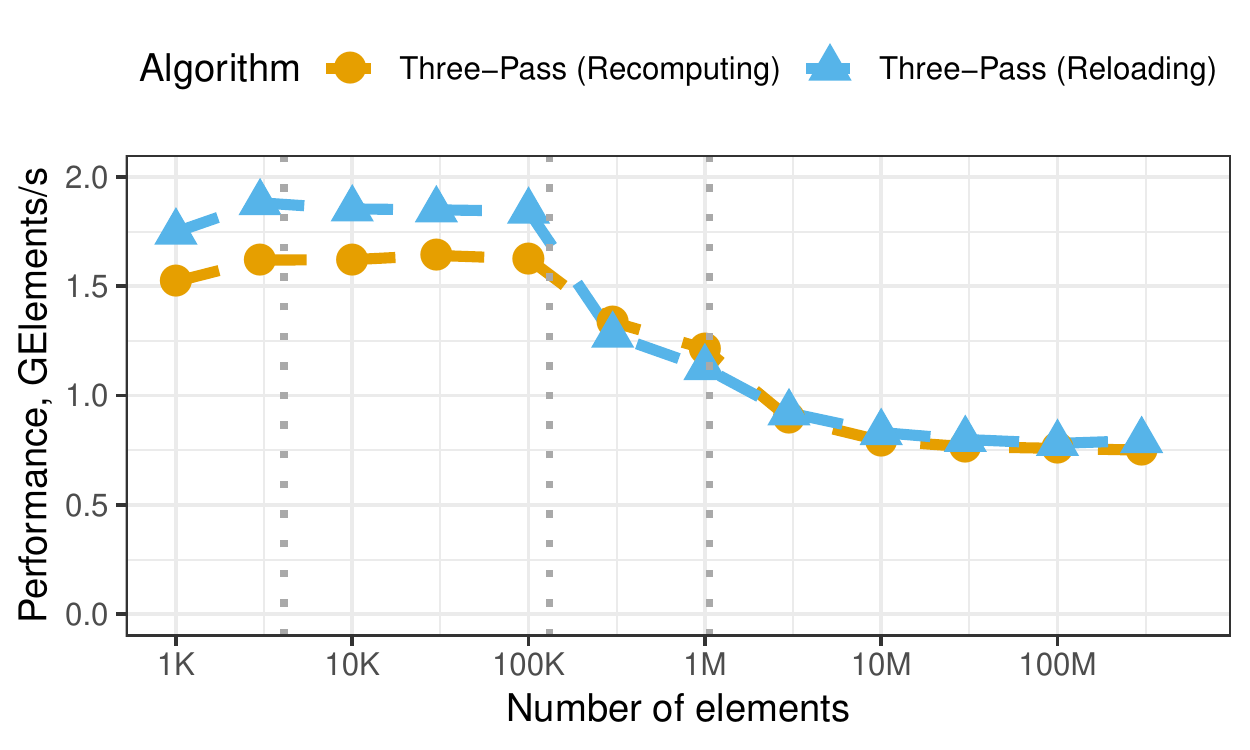}
    \caption{Performance comparison of the Softmax algorithms~\ref{algo:three_pass} and~\ref{algo:three_pass_with_reloading} in the AVX2 implementations on the Skylake-X system. Gray dotted lines denote boundaries of level-1, level-2, and level-3 caches.}
    \label{fig:avx2-threepass}
\end{figure}

The AVX2 implementation of the same Three-Pass softmax Algorithm~\ref{algo:three_pass} and Algorithm~\ref{algo:three_pass_with_reloading} is illustrated in Fig.~\ref{fig:avx2-threepass} and demonstrates similar trends. As the working set increases, the $13-16\%$ speedup from reloading of exponential computations goes down, and eventually levels off at $3-6\%$ for large arrays.

The small difference between recomputing and reloading of exponential computations on Fig.~\ref{fig:avx512-threepass} and Fig.~\ref{fig:avx2-threepass} suggests that despite the expensive exponential function, softmax computation might be memory-bound for large arrays. To directly test this hypothesis, we decompose the Algorithms~\ref{algo:three_pass} and~\ref{algo:three_pass_with_reloading} into individual memory passes, and compare measured bandwidth to STREAM benchmarks~\cite{STREAM}.

\begin{figure}[H]
    \includegraphics[width=\linewidth]{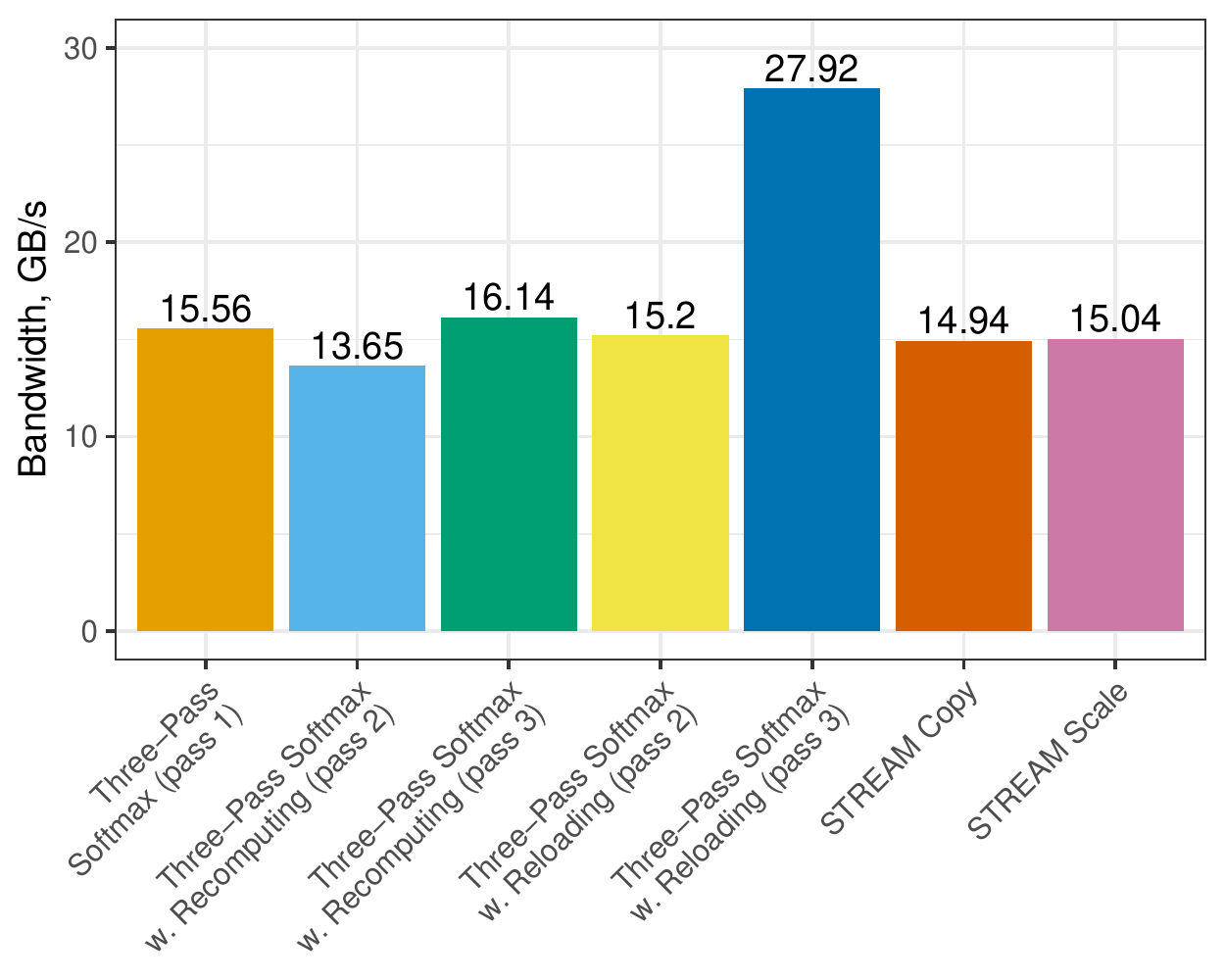}
    \caption{Measured memory bandwidth on the Skylake-X system in the three passes of the Softmax algorithms~\ref{algo:three_pass} and~\ref{algo:three_pass_with_reloading}, and in the STREAM benchmark. Both the softmax implementations and the STREAM benchmark use AVX512 instructions.}
    \label{fig:avx512-stream}
\end{figure}

\begin{figure}[H]
    \includegraphics[width=\linewidth]{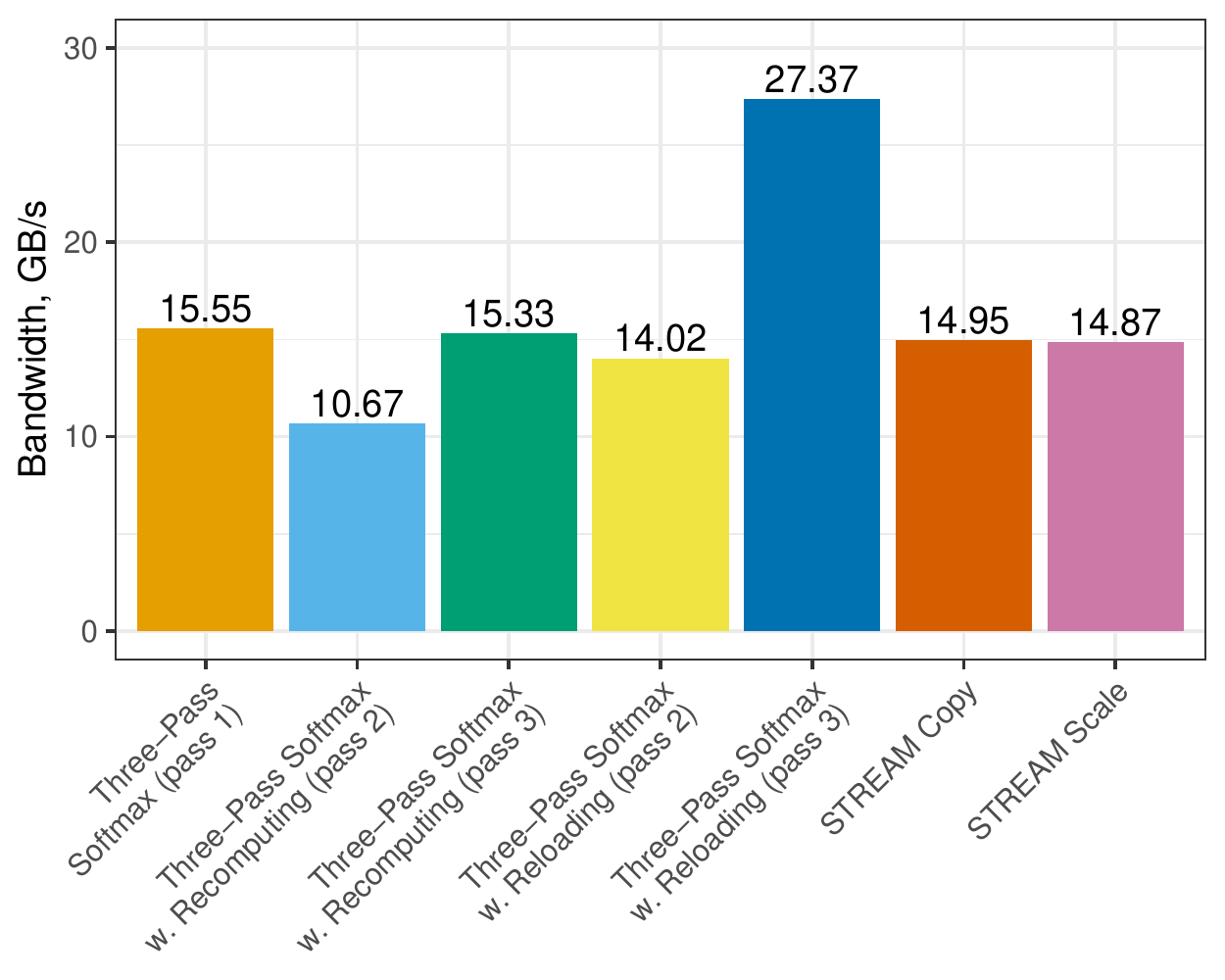}
    \caption{Measured memory bandwidth on the Skylake-X system in the three passes of the Softmax algorithms~\ref{algo:three_pass} and~\ref{algo:three_pass_with_reloading}, and in the STREAM benchmark. Both the softmax implementations and the STREAM benchmark use AVX2 instructions.}
    \label{fig:avx2-stream}
\end{figure}

Fig.~\ref{fig:avx512-stream} and~\ref{fig:avx2-stream} illustrate the memory bandwidth in each pass of the Three-Pass softmax Algorithms~\ref{algo:three_pass} and~\ref{algo:three_pass_with_reloading}, as well as Copy and Scale STREAM benchmarks~\cite{STREAM}. As recommended in STREAM documentation, we set the array size to four times the size of last-level cache (8650752 single-precision elements for Softmax, and 4325376 double-precision elements for STREAM). The first softmax pass (max-reduction) is the same in both versions of the Three-Pass algorithm, and thus presented only once. This pass reads one input array, and doesn't have a direct equivalent in STREAM, but achieves similar bandwidth to STREAM Copy and Scale benchmarks, which both read one array and write one array. The second pass in Algorithm~\ref{algo:three_pass} reads one array, computes exponentials on the inputs, and accumulates them. It achieves $91\%$ of STREAM Copy bandwidth in AVX512 version and $71\%$ in AVX2 version. The second pass Algorithm~\ref{algo:three_pass_with_reloading} is similar, but additionally stores computed exponents into the output array. Although it achieves higher bandwidth than the second pass in Algorithm~\ref{algo:three_pass}, it takes substantially longer to complete; the higher bandwidth is due to doubling the number of transferred bytes with a less than proportional increase in run time. The third pass of Algorithm~\ref{algo:three_pass} reads one array, computes exponentials on the inputs, scales them, and writes results to another array. This pass does the same number of memory operations as STREAM Scale benchmark, but substantially more computational operations. Yet, our auto-tuned implementations exceed the performance of STREAM Scale benchmark in both the AVX512 and the AVX2 versions. The third pass of the Algorithm~\ref{algo:three_pass_with_reloading} is an in-place variant of STREAM Scale benchmark. The processor clearly favors in-place operation: it is $86\%$ faster than STREAM Scale with AVX512, and $84\%$ faster with AVX2.

To summarize, passes 1 and 3 of the Algorithm with Recomputing~\ref{algo:three_pass} demonstrate similar memory performance to STREAM benchmark, and pass 2 in AVX512 implementation is not far behind. Passes 1 and 2 of the Algorithm~\ref{algo:three_pass_with_reloading} with Reloading similarly perform close to STREAM bandwidth, and pass 3 is significantly faster than STREAM Scale benchmark. These results confirm that performance of Three-Pass softmax algorithms is limited by achievable memory bandwidth, and suggest that \textbf{softmax computation can be further accelerated only through reducing the number of memory operations}.

\subsection{The Two-Pass Algorithm}

\label{sec:two_pass_evaluation}

\begin{figure}[H]
    \includegraphics[width=\linewidth]{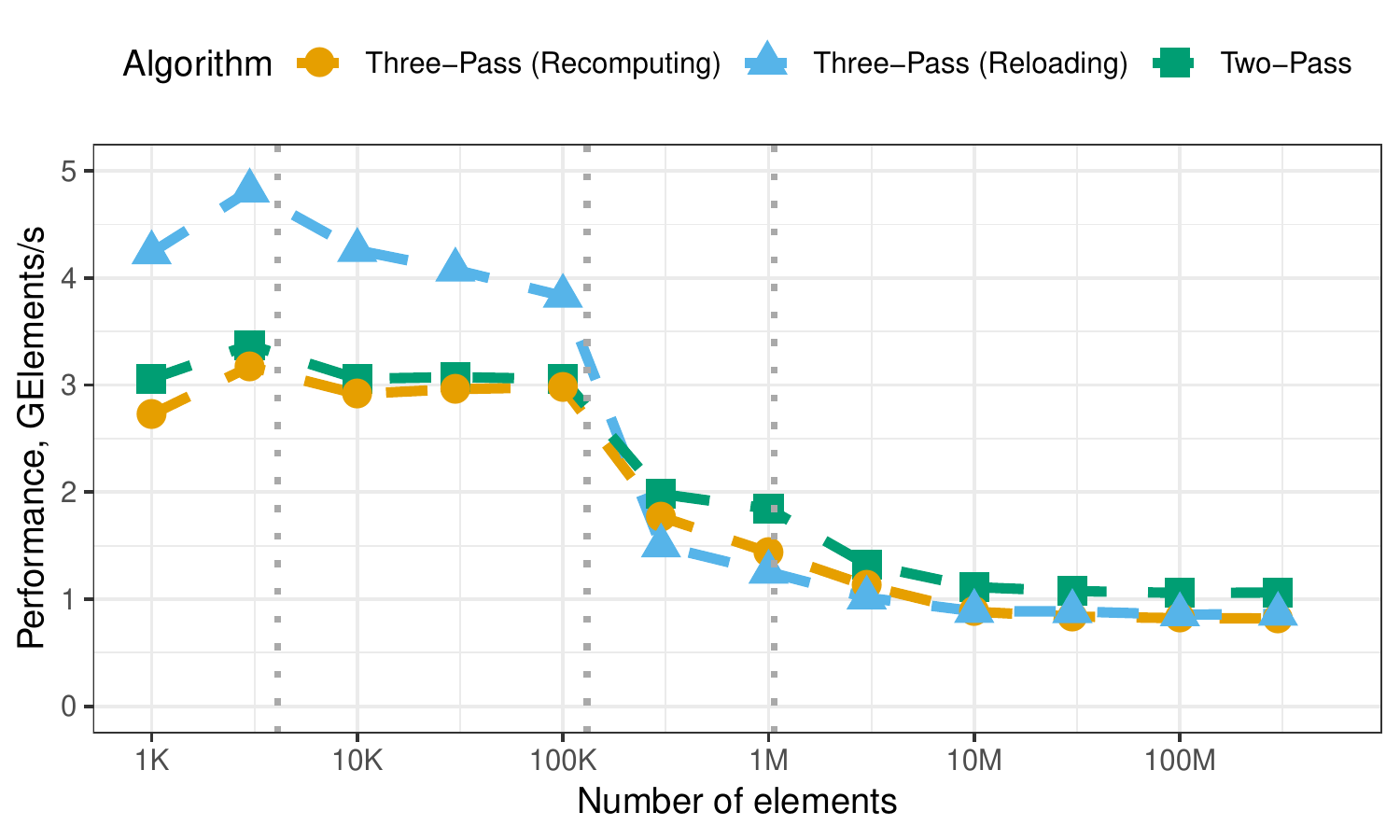}
    \caption{Performance comparison of the Algorithms~\ref{algo:three_pass}, \ref{algo:three_pass_with_reloading}, and~\ref{algo:two-pass} in the AVX512 implementations. Gray dotted lines denote boundaries of level-1, level-2, and level-3 caches.}
    \label{fig:avx512}
\end{figure}

On Fig.~\ref{fig:avx512} we present the performance of the Two-Pass softmax algorithm in comparison with the two versions of the Three-Pass algorithm in AVX512 implementations. On out-of-cache working sets the proposed Two-Pass softmax algorithm outperforms Three-Pass algorithms by $18\%-28\%$.

\begin{figure}[H]
    \includegraphics[width=\linewidth]{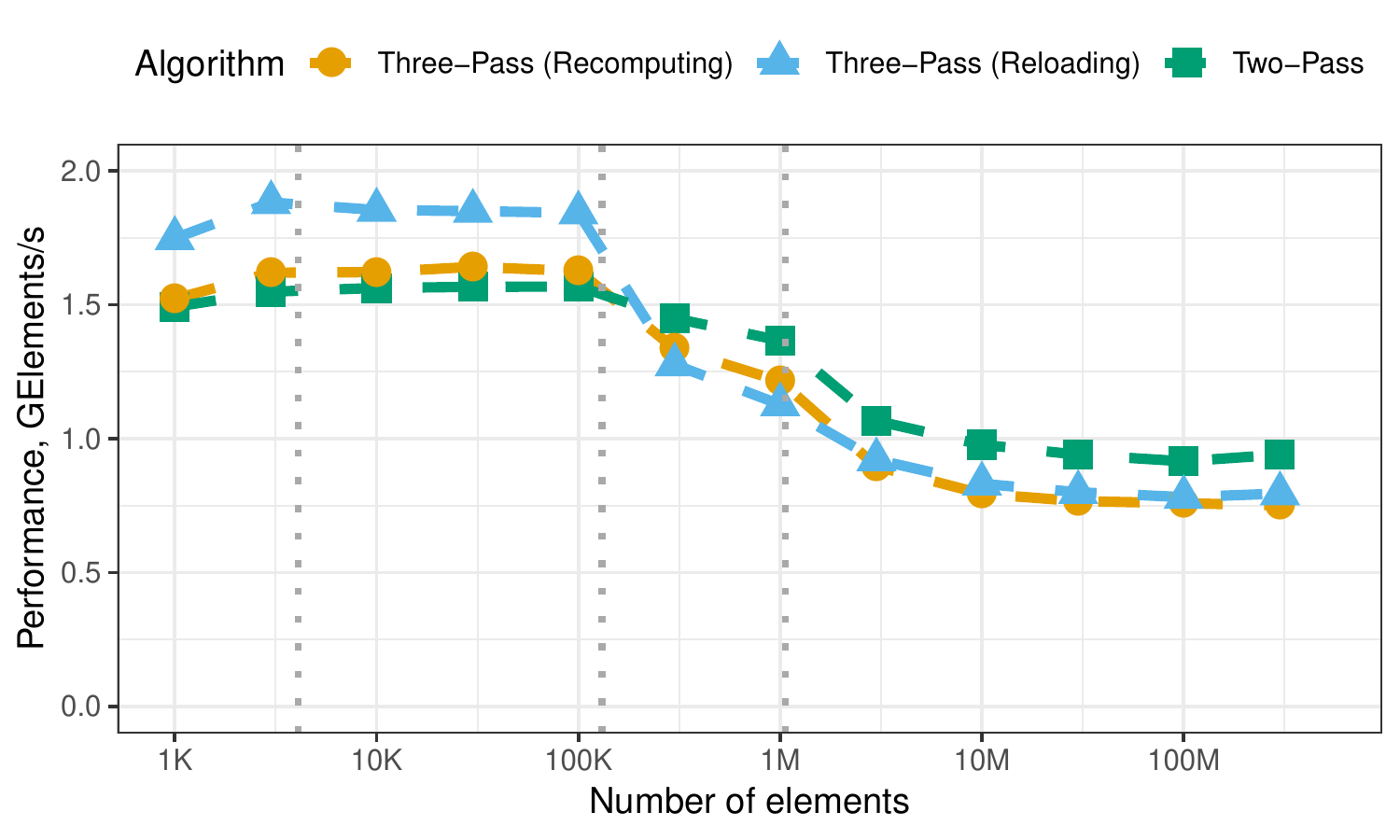}
    \caption{Performance comparison of the Algorithms~\ref{algo:three_pass}, \ref{algo:three_pass_with_reloading}, and~\ref{algo:two-pass} in the AVX2 implementations. Gray dotted lines denote boundaries of level-1, level-2, and level-3 caches.}
    \label{fig:avx2}
\end{figure}

Fig.~\ref{fig:avx2} similarly compares performance of the Two-Pass and the Three-Pass algorithms in the AVX2 implementations. Here, the Two-Pass algorithm outperforms Three-Pass algorithm with Reloading of exponential computations by $16\%-18\%$ on out-of-cache workloads. The smaller speedups, compared to AVX512 implementation, are explained by relatively higher cost of recomputing exponentials in AVX2 compared to AVX512. If we compare to the Three-Pass Algorithm~\ref{algo:three_pass}, which similarly recomputed exponentials, the Two-Pass algorithm wins by $19-25\%$.

\begin{figure*}
    \includegraphics[width=\linewidth]{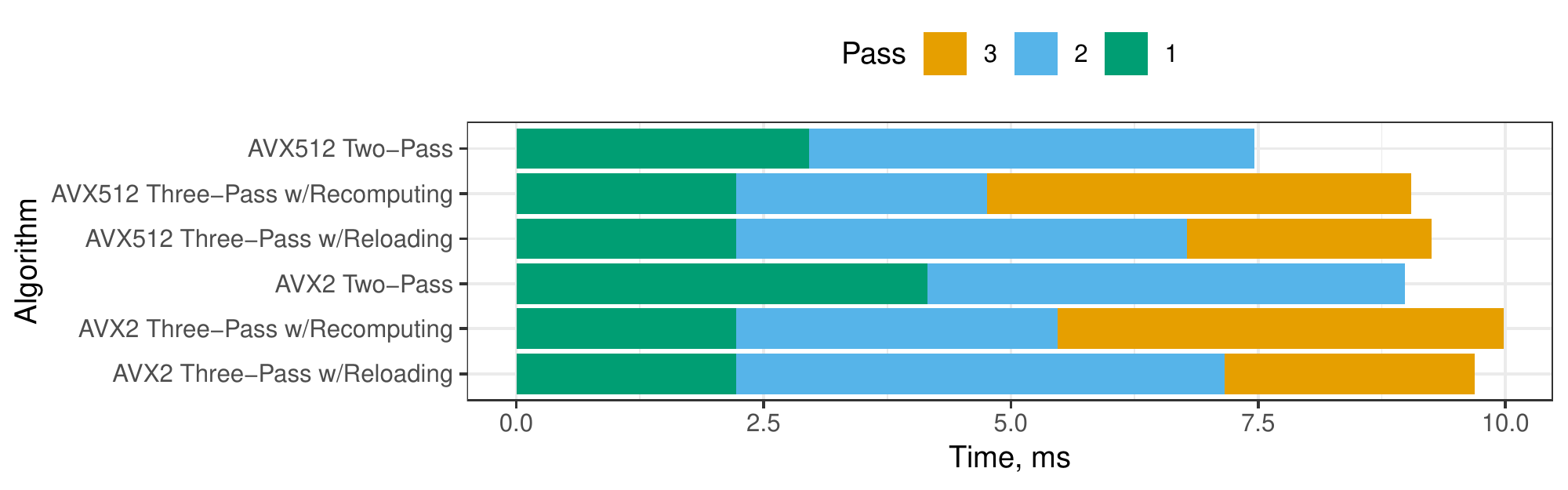}
    \caption{Absolute runtime of the passes in the Algorithms~\ref{algo:three_pass}, \ref{algo:three_pass_with_reloading}, and~\ref{algo:two-pass} in both the AVX2 and the AVX512 implementations. The algorithms were evaluated on arrays of 8,650,752 single-precision elements on the Skylake-X system.}
    \label{fig:runtime}
\end{figure*}

On Fig.\ref{fig:runtime} we decompose the absolute run-time for the three algorithms and two SIMD instruction sets into individual memory passes and offers insight into the origin of performance improvements with the Two-Pass algorithm. The two passes of the Two-Pass softmax algorithm have similar, but slightly higher absolute run-time to the last two passes of the Three-Pass softmax algorithm with recomputation of re-computation of exponential function, which share the same memory access pattern. The slightly higher run-time in the passes of the Two-Pass algorithm can be explained by larger number of operations needed for accumulation on the $(m, n)$ representation compared to just accumulating scalar floating-point values.

\subsection{Multi-Threaded Performance}

\label{sec:multithreaded_evaluation}

The benchmarks in Sec.~\ref{sec:three_pass_evaluation} and Sec.~\ref{sec:two_pass_evaluation} presented performance in single-threaded softmax computation, and demonstrated that on HPC-class systems softmax saturates memory bandwidth even when running on a single core. Utilizing multiple cores increases available computational resources faster than achievable memory bandwidth, and therefore increases the advantage of the bandwidth-saving Two-Pass softmax algorithm. To quantify this advantage, we fix the size of the array at 4 times the last-level cache size~\cite{STREAM}, and vary the number of threads from 1 to 6 (number of cores in the system) to 12 (number of logical processors, including hyperthreads, in the system).

\begin{figure}[H]
    \includegraphics[width=\linewidth]{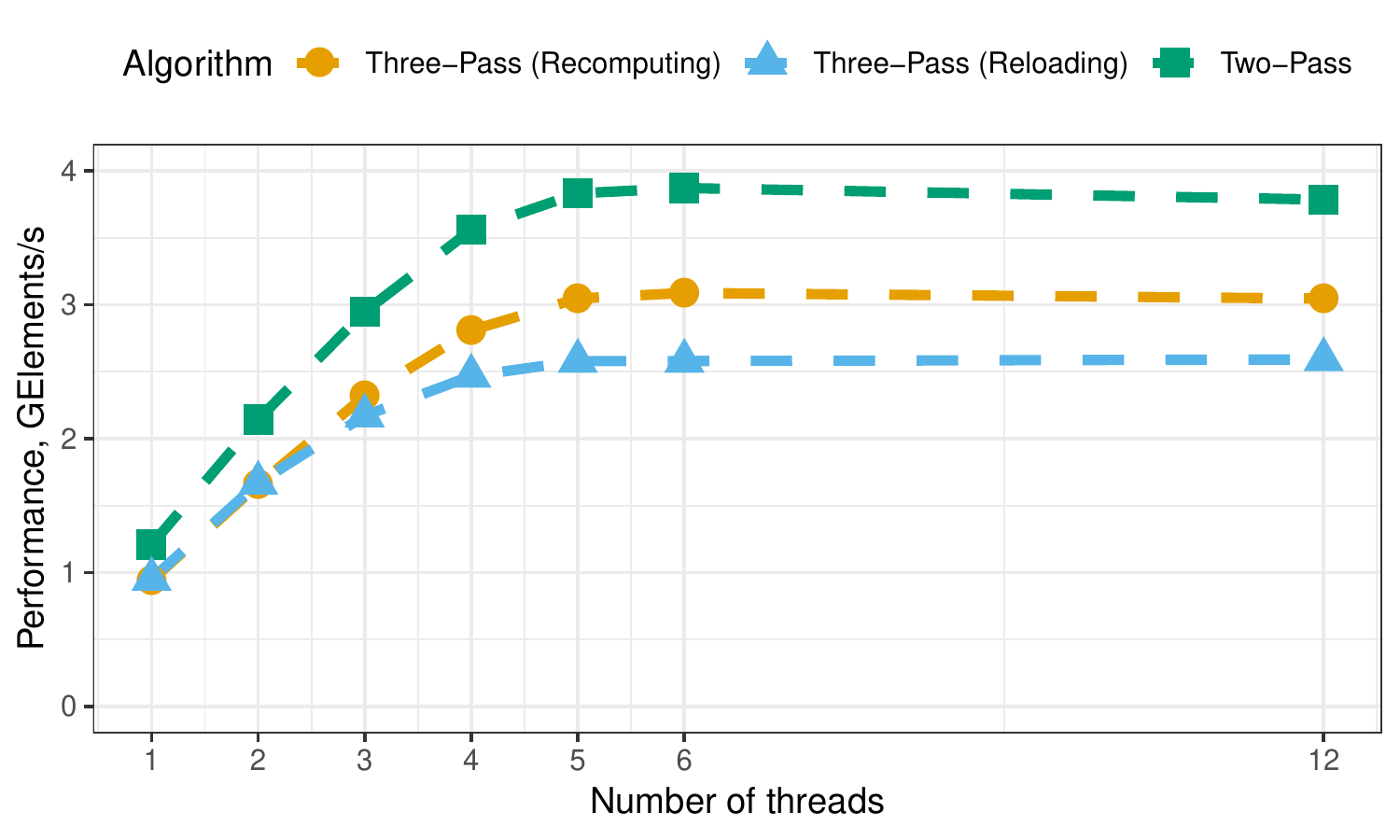}
    \caption{Weak scaling of the softmax algorithms in the AVX512 implementations on the Skylake-X system.}
    \label{fig:avx512-scaling}
\end{figure}

Fig.~\ref{fig:avx512-scaling} illustrates weak multi-core scaling of the AVX512 implementations. As the number of threads grows, the advantage of the Two-Pass over Three-Pass algorithms remains unchanged at $25-28\%$. Interestingly, the reloading variant of the Three-Pass algorithm scales worse than the recomputing variant, and the recomputing Algorithm~\ref{algo:three_pass} outperforms the reloading Algorithm~\ref{algo:three_pass_with_reloading} when at least 3 cores are being utilized.

\begin{figure}[H]
    \includegraphics[width=\linewidth]{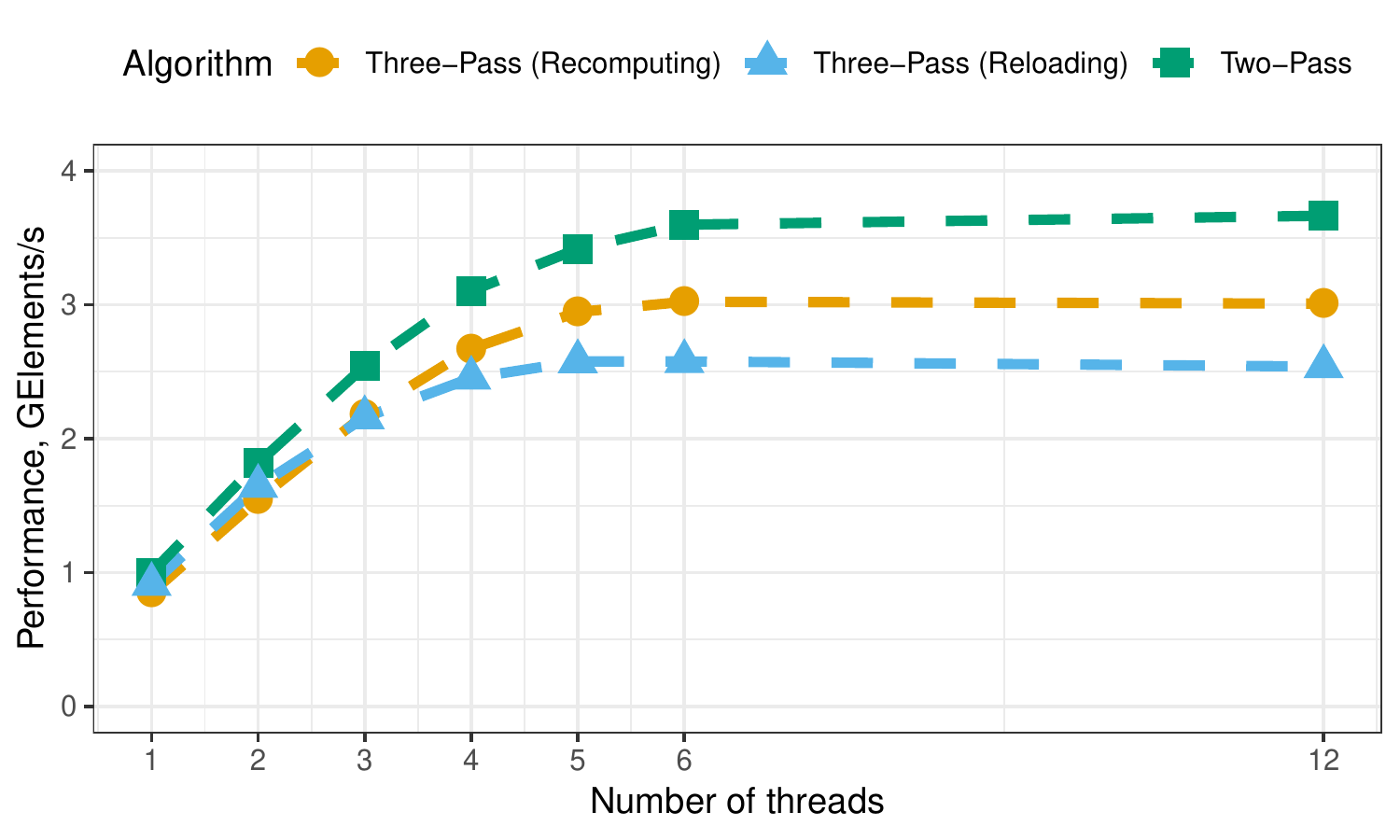}
    \caption{Weak scaling of the softmax algorithms in the AVX2 implementations on the Skylake-X system.}
    \label{fig:avx2-scaling}
\end{figure}

Fig.~\ref{fig:avx2-scaling} similarly illustrates weak multi-core scaling of the AVX2 implementations. The advantage of the Two-Pass over Three-Pass algorithms grows from $9\%$ on a single core to $19\%$ when utilizing all 6 cores to $22\%$ when also using hyperthreads. 

\subsection{Comparison with Intel DNNL}

\label{sec:dnnl_comparison}

The results in Sec.~\ref{sec:two_pass_evaluation}-\ref{sec:multithreaded_evaluation} demonstrate that on out-of-cache inputs the Two-Pass softmax algorithm outperforms the Three-Pass softmax algorithms in our implementation, but leaves out the question of whether our implementations are competitive with state-of-the-art. To demonstrate the absolute effectiveness of the Two-Pass algorithm, we compared our implementations of the three softmax algorithm to the softmax primitive in Intel Deep Neural Network Library (DNNL) version 1.1.1.

Intel DNNL implements the Three-Pass softmax Algorithm~\ref{algo:three_pass_with_reloading} with reloading of computed exponentials. It includes implementations for SSE4.1, AVX, and AVX512 instruction sets, and automatically dispatch to AVX512 implementation on the Skylake-X processor. Unlike our implementations, Intel DNNL generates implementation in runtime using Just-in-Time (JIT) technology. JIT code generation potentially allows adaptation of implementation to parameters of a particular softmax operator (e.g. number of channels).

\begin{figure*}
    \includegraphics[width=\linewidth]{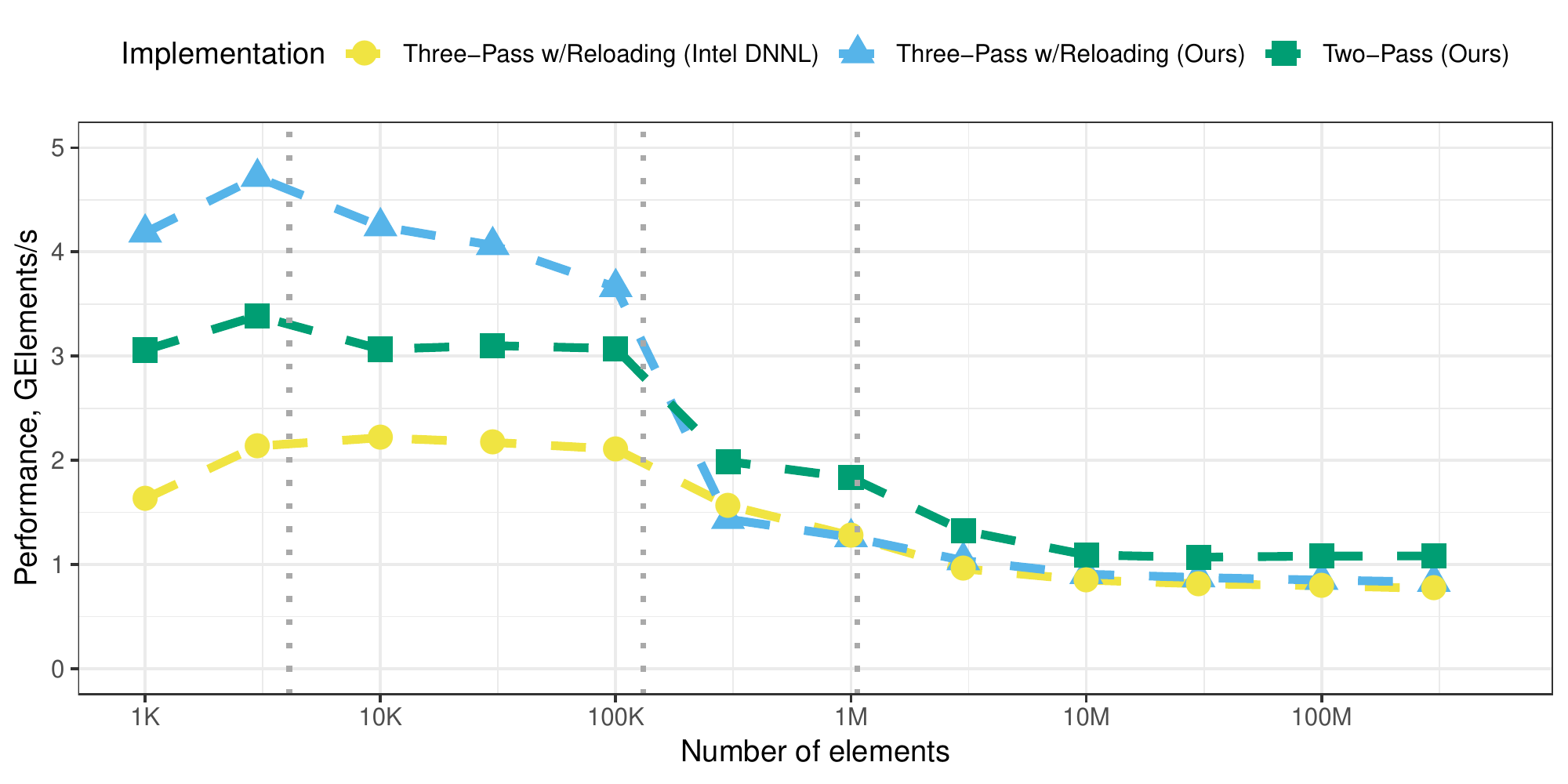}
    \caption{Performance comparison of our implementation of Algorithms~\ref{algo:three_pass}, \ref{algo:three_pass_with_reloading}, and~\ref{algo:two-pass}, with the softmax implementation in Intel DNNL library. Gray dotted lines denote boundaries of level-1, level-2, and level-3 caches.}
    \label{fig:skylake-dnnl}
\end{figure*}

Fig.~\ref{fig:skylake-dnnl} presents the comparison between two implementations (Ours and DNNL) of the Three-Pass algorithm with reloading of exponentials, and the Two-Pass softmax algorithm in our implementation. For the Three-Pass algorithm with reloading, our implementation ourperforms Intel DNNL for the majority of problem sizes. Its advantage is particularly high -- over 2X -- when data fits into L1, diminish to $72-94\%$ within L2, and levels off at $7-8\%$ for out-of-cache problem sizes. As the implementation in Intel DNNL is less efficient than ours, our Two-Pass softmax algorithm outperforms DNNL softmax primitive on all problem sizes: it is $28-41\%$ faster on out-of-cache problem sizes, and up to $87\%$ when input fits into L1 cache.

\subsection{Validation on Alternative Processors}

\label{sec:alternative_processors}

The results in Sec.~\ref{sec:three_pass_evaluation}-\ref{sec:dnnl_comparison} were all collected on the Xeon W-2135 processor with the Intel Skylake-X microarchitecture, which prompts a question as to whether the advantage of the Two-Pass softmax algorithm is restricted to a specific type of processor. To demonstrate that the Two-Pass algorithm generalize to other types of processors, we replicated results of Sec.~\ref{sec:two_pass_evaluation} on Xeon E5-2696 v4 processor with Intel Broadwell microarchitecture and Ryzen 9 3900X with AMD Zen 2 microarchitecture. Both of these processors support the AVX2, but not the AVX512 instruction set, and have different cache hierarchy parameters than the Intel Skylake-X system.

\begin{figure}[H]
    \includegraphics[width=\linewidth]{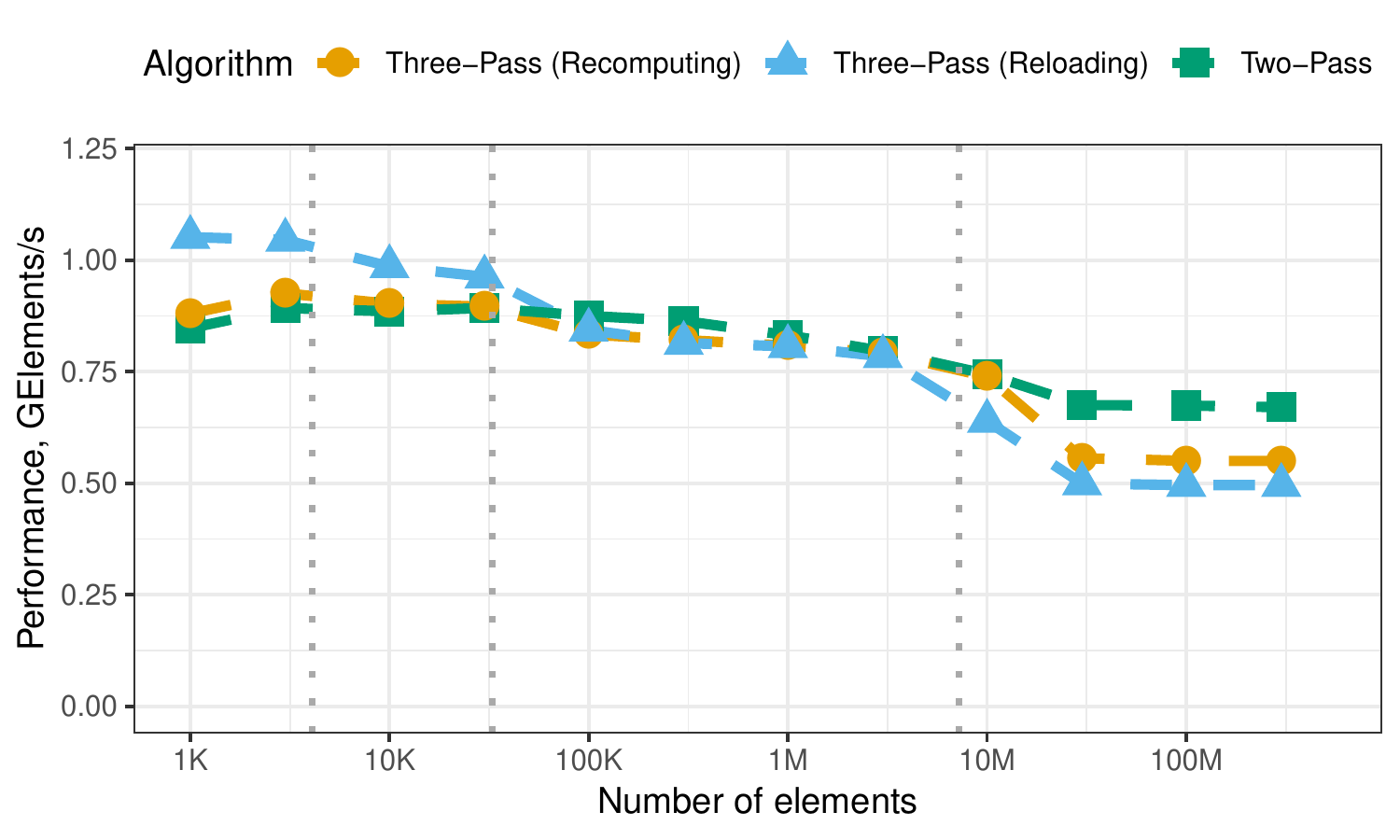}
    \caption{Performance comparison of Algorithms~\ref{algo:three_pass}, \ref{algo:three_pass_with_reloading}, and~\ref{algo:two-pass} on an Intel Broadwell-based system. Gray dotted lines denote boundaries of level-1, level-2, and level-3 caches.}
    \label{fig:broadwell}
\end{figure}

Fig.~\ref{fig:broadwell} presents performance of the three softmax algorithms on the Intel Broadwell system. Although the Two-Pass softmax algorithm demonstrates inferior performance on problems which fit into L2 cache, it gets competitive with the Three-Pass softmax algorithms on L3-sizes problems, and outperforms them by $21-23\%$ on out-of-cache problems.

\begin{figure}[H]
    \includegraphics[width=\linewidth]{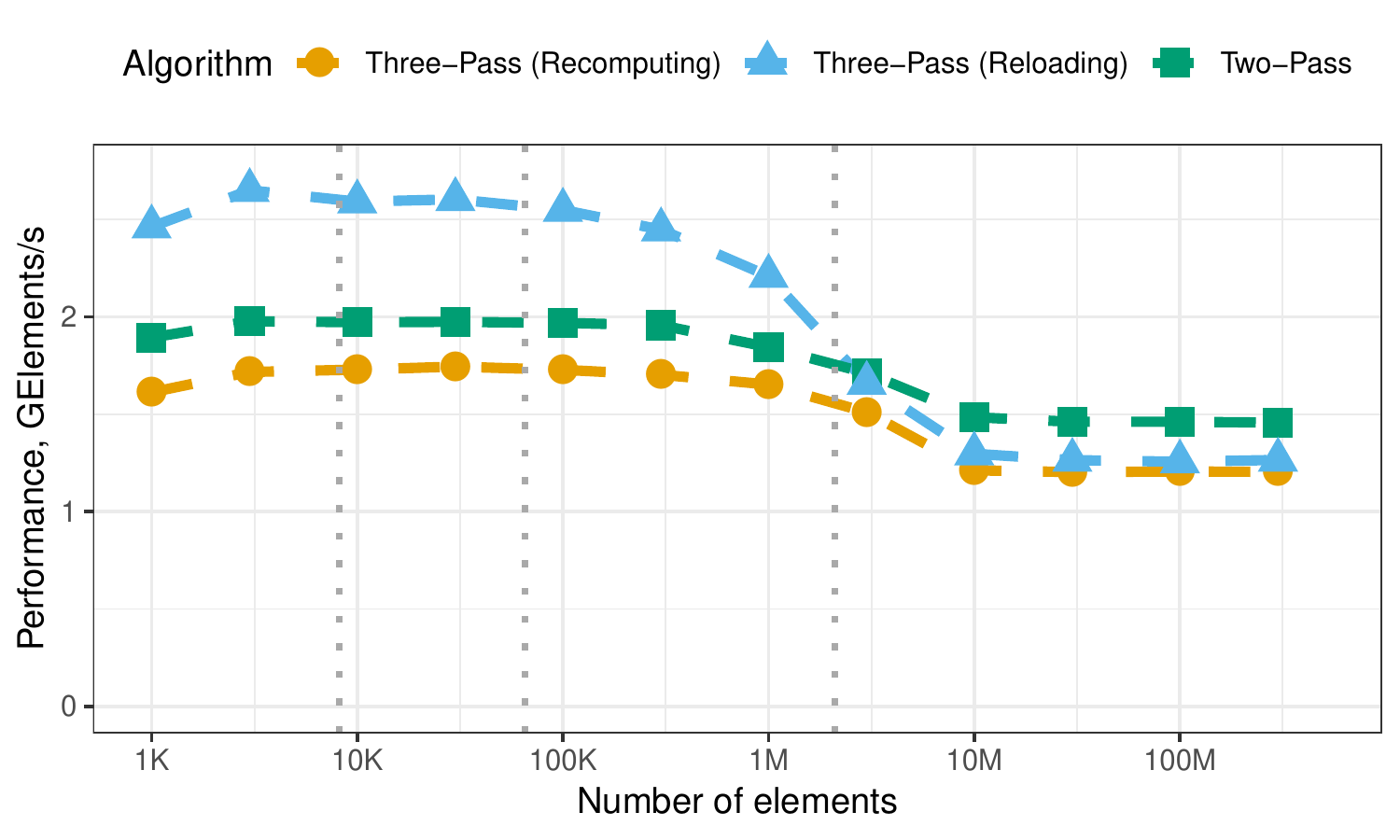}
    \caption{Performance comparison of Algorithms~\ref{algo:three_pass}, \ref{algo:three_pass_with_reloading}, and~\ref{algo:two-pass} on a Ryzen 9 3900X system. Gray dotted lines denote boundaries of level-1, level-2, and level-3 caches.}
    \label{fig:zen2}
\end{figure}

Fig.~\ref{fig:zen2} shows a similar picture on AMD Zen 2 microarchitecture. Here, the Three-Pass algorithms deliver superior performance as long as data fits into L3 cache, but loose $14-16\%$ to the Two-Pass algorithm when the data exceeds cache size.

\section{Conclusion}

We presented a novel Two-Pass algorithm for softmax computation and demonstrated that the new Two-Pass algorithm is up to 28\% faster than the traditional Three-Pass algorithm on large input vectors. The algorithm, however, offers no advantage over reloading variant of the Three-Pass algorithm when the data fits into the processor's cache.

This study focused on performance on a CPU, but the algorithm has great potential for GPU and hardware AI accelerators. These platforms further shift the balance between compute and memory performance towards expensive memory and cheap floating-point operations, and would favor reduced memory intensity of the presented Two-Pass softmax algorithm.

To foster reproducibility, we released an open-source implementation of the new Two-Pass Softmax algorithm and other experiments in this paper as a part of XNNPACK library at \href{http://www.github.com/google/XNNPACK}{GitHub.com/google/XNNPACK}.

\section*{Acknowledgements}

We thank Matthias Grundmann, Sergey Ioffe, Juhyun Lee, Karthik Raveendran, and Richard Vuduc for helpful feedback and discussion, and Vijay Thakkar for sharing access to the Ryzen 9 3900X system.

\bibliographystyle{IEEEtran}
\bibliography{softmax}

\end{document}